\documentclass[10pt,a4paper]{article}
\usepackage{graphicx}
\usepackage{epstopdf}
\usepackage{dcolumn}
\usepackage{authblk}
\usepackage[usenames]{color}
\usepackage{bm}
\usepackage{amsmath}
\usepackage{amssymb}
\usepackage[normalem]{ulem}
\usepackage[margin=1in,footskip=0.25in]{geometry}

\def\GK{$\overline{\Gamma {\rm K}}$\ }
\def\GM{$\overline{\Gamma {\rm M}}$\ }

\newcommand{\degree}{\ensuremath{^\circ}}

\newcommand{\bra}[1]{\langle\,#1\,|}
\newcommand{\ket}[1]{|\,#1\,\rangle}

\renewcommand{\figurename}{\textbf{Figure}}

\begin{document}

\title{Synchrotron radiation induced magnetization in magnetically-doped and pristine topological insulators}

\author[1]{A.~M.~Shikin}
\author[1]{D.~M.~Sostina}
\author[1]{A.~A.~Rybkina}
\author[1]{V.~Yu.~Voroshnin}
\author[1]{I.~I.~Klimovskikh}
\author[1]{A.~G.~Rybkin}
\author[1]{D.~A.~Estyunin}
\author[1,2,3]{K.~A.~Kokh}
\author[1,2,4]{O.~E.~Tereshchenko}
\author[5]{L.~Petaccia}
\author[5]{G.~Di~Santo}
\author[6,7,8]{P.~N.~Skirdkov}
\author[6,7,8]{K.~A.~Zvezdin}
\author[6,7,8]{A.~K.~Zvezdin}
\author[9]{A.~Kimura}
\author[1,10,11,12]{E.~V.~Chulkov}
\author[10,11,13]{E.~E.~Krasovskii}

\affil[1]{Saint Petersburg State University, Saint Petersburg, 198504 Russia}
\affil[2]{Novosibirsk State University, Novosibirsk, 630090 Russia}
\affil[3]{V.S. Sobolev Institute of Geology and Mineralogy, Novosibirsk, 630090 Russia}
\affil[4]{A.V. Rzhanov Institute of Semiconductor Physics, Novosibirsk, 630090 Russia}
\affil[5]{Elettra Sincrotrone Trieste, Strada Statale 14 km 163.5, 34149 Trieste, Italy}
\affil[6]{Moscow Institute of Physics and Technology, Institutskiy per. 9, 141700 Dolgoprudny, Russia}
\affil[7]{A.M. Prokhorov General Physics Institute, Russian Academy of Sciences, Vavilova 38, 119991 Moscow, Russia}
\affil[8]{Russian Quantum Center, Novaya St. 100, 143025 Skolkovo, Moscow Region, Russia}
\affil[9]{Graduate School of Science, Hiroshima University,1-3-1 Kagamiyama, Higashi-Hiroshima 739-8526, Japan}
\affil[10]{Departamento de F\'{i}sica de Materiales, Facultad de Ciencias Qu\'{i}micas, UPV/EHU, San Sebasti\'{a}n/Donostia, 20080 Basque Country, Spain}
\affil[11]{Donostia International Physics Center (DIPC), San Sebasti\'{a}n/Donostia, 20018 Basque Country, Spain}
\affil[12]{Centro de Fisica de Materiales CFM - MPC and Centro Mixto CSIC-UPV/EHU, San Sebasti\'{a}n/Donostia, 20080 Basque Country, Spain}
\affil[13]{IKERBASQUE, Basque Foundation for Science, 48013 Bilbao, Spain}

\maketitle

\begin{abstract}
Quantum mechanics postulates that any measurement influences the state
of the investigated system.  Here, by means of angle-, spin-, and
time-resolved photoemission experiments and ab initio calculations we
demonstrate how non-equal depopulation of the Dirac cone (DC) states
with opposite momenta in V-doped and pristine topological insulators
(TIs) created by a photoexcitation by linearly polarized 
synchrotron radiation (SR) is followed by the hole-generated uncompensated spin accumulation and 
the SR-induced magnetization via the spin-torque
effect. We show that the photoexcitation of the DC is asymmetric, that it varies with the 
photon energy, and that it practically does not change during the 
relaxation.  We find a relation between the photoexcitation asymmetry, the generated 
spin accumulation and the induced spin polarization of the DC and V 3$d$ states. Experimentally the
SR-generated in-plane and out-of-plane magnetization is confirmed by
the $k_{\parallel}$-shift of the DC position and by the splitting of the
states at the Dirac point even above the Curie temperature. Theoretical predictions and estimations of the measurable physical quantities substantiate the experimental results.
\end{abstract}

The photoexcitation by laser or synchrotron radiation 
is accompanied by a depopulation of the initial states, which influences the 
electronic structure observed in photoemission (PE) 
measurements. In materials with helical spin structure (for instance, topological 
insulators (TIs) \cite{Hasan2010RevMPhys, Hsieh2009, Zhang2009, Moore2009}) the
imbalance in photoexcitation of the DC states with opposite momenta created, for 
instance, by circularly polarized laser or SR can be effectively used for the 
generation of the surface spin-polarized currents that depend on the helicity of the radiation polarization 
\cite{Hosur2011, Junck2013, McIverJ2012n, Ogawa2016, Shikin2016, Kastl2015}. 
Similar to the case of an electric field applied in the surface plane 
\cite{LiC2014n, Mellnik2014, Fan2014}, this can induce a magnetization in ferromagnetic TIs \cite{Ogawa2016, Shikin2016APL}. 
The induced magnetic moment opens a gap at the Dirac point (DP) due to Time Reversal 
Symmetry (TRS) breaking providing a platform for the realization of unique 
quantum phenomena such as quantized magneto-electric effect 
\cite{Qi-2008PRB, Nomura-2011PRL, Wang-2015PRB} 
and quantum anomalous Hall effect \cite{Chang-2013Sci, Chang-2015NatMat, Yu2010} at elevated
temperatures under light excitation. Although, the possibility of induced (and 
controlled) magnetization by linearly polarized SR has not been studied yet, the generation of 
spin-polarized current was noted recently 
\cite{McIverJ2012n, Ogawa2016, Shikin2016, Hasan2010RevMPhys}. The present 
work aims to investigate a possibility of such SR-induced in-plane and out-of-plane
magnetization in FM-doped and pristine TIs by linearly polarized SR. 
We relate this phenomenon to an asymmetry in the depopulation 
of the DC states with opposite momenta, which leads to a hole-generated uncompensated spin accumulation.
The possibility of a long-living electron-hole separation between the excited electrons and generated holes 
(related to a reduced electron-phonon interaction at the surface 
\cite{Neupane2015,  Crepaldi2013}) is confirmed by a series of time-resolved laser experiments 
(see, for instance, \cite{Hajlaoui2014, Neupane2015, Crepaldi2013, Ishida2015}). 
The holes generated at the topological surface states (TSSs) are compensated by 
a drift of electrons from the TSSs out of the beam spot via long-time two-dimensional 
relaxation process \cite{Neupane2015}. In the case of a different probability of the photoexcitation of electrons from the DC states with opposite momenta, the  uncompensated spin accumulation and the zero-bias spin-polarized photocurrent occur, as in Refs.~\cite{Ogawa2016, Shikin2016, Shikin2016APL}, which can lead to induced magnetization similar to the case discussed in Ref.~\cite{Chernyshov2009} 
for the (Ga,Mn)As ferromagnetic semiconductor.

The issue of SR-generated spin accumulation is strongly related to the TSS spin texture 
and to the asymmetry of the photoexcitation of the TSSs with opposite spin 
orientation. In Refs. \cite{Zhang2013PRL, Cao2013}, it was shown that a total spin texture 
of TSSs includes not only contributions of the 
$p_z$-orbitals, but also of the $p_x$- and $p_y$-orbitals related to the radial and 
tangential components in the spin-orbital texture, which significantly modifies the spin texture of the TSSs probed 
by photoemission. Total spin structure in PE spectra depends on 
the sum of all contributions determined by optical selection rules, 
quantum interference and SR incident angle \cite{Zhu2013, Zhu2014, Seibel2016}. Due to variation of $k_z$ with photon
energy all $p_{x, y, z}$ components ``oscillate'' with different phases. The non-trivial 
character of TSS spin texture is confirmed by spin-resolved photoemission both via theoretical analysis 
\cite{Jozwiak2013, Zhu2013, Zhu2014, Park2012, Jozwiak2011, Cao2013, Seibel2016} and via experimental measurements
\cite{Cao2013, Zhu2013, Zhu2014, Seibel2016} using different polarization of SR
\cite{Jozwiak2013, Neupane2013, Zhu2013, Zhu2014} and manifests itself
in significant modification of the spin polarization of photoelectrons
with photon energy \cite{Jozwiak2013, Zhu2013, Zhu2014, Park2012}. The
oblique incidence of SR breaks the symmetry of the angular distribution of the photocurrent, and the
different photoemission intensity at $k_{\parallel}$ and $-k_{\parallel}$ 
\cite{Zhu2013, Cao2013} can be a source of the SR-generated uncompensated 
spin accumulation and an induced magnetization. This problem is especially 
important for magnetically-doped TIs because the DP gap opening due to the TRS breaking, its value and 
origin (magnetization- or hybridization-derived) are being actively discussed, see, for instance, Ref.~\cite{Xu2012}. 
However, without a proper analysis of the influence of the non-equal depopulation of the TSSs on a
possible induced magnetization during ARPES and SARPES measurements such questions cannot be answered.

In the present work we study the PE intensity asymmetry of the DC states and connect it with the SR-generated 
asymmetric $k_\parallel$-distribution of the holes and the resulting spin accumulation leading 
a local induced magnetization via spin-torque effect. The first part of the 
work aims to clarify how the photoexcitation by a linearly polarized SR influences 
the DC states intensity distribution in the ARPES intensity 
maps and how it varies with photon energy. In the second part we analyze how the
 imbalance in the depopulation of the TSSs in V-doped TI and the 
corresponding SR-generated uncompensated spin accumulation result in an induced 
in-plane and out-of-plane polarization of TSSs and the V 3$d$-ions. Then, 
the laser pump-probe experiment allows us to conclude that the imbalance 
in the depopulation of the TSSs is practically not changed during the relaxation, 
and the SR-induced magnetization can be estimated in rough
approximation using the TSS intensity asymmetry in PE spectra.

\section*{Asymmetry in the intensity of the DC states vs photon energy}

In this work we study a series of pristine and V-doped TIs with fractional 
stoichiometry based on Bi$_2$Te$_2$Se with inclusion of different Sb-concentrations. 
They have a wide insulating energy gap with the DP inside the bulk gap \cite{Miyamoto2012, Shikin2014} and an 
enhanced surface contribution to the spin transport \cite{Tang2013}, which is important for spintronics applications.

\begin{figure}[t]
\centering
\includegraphics[width=0.81\textwidth]{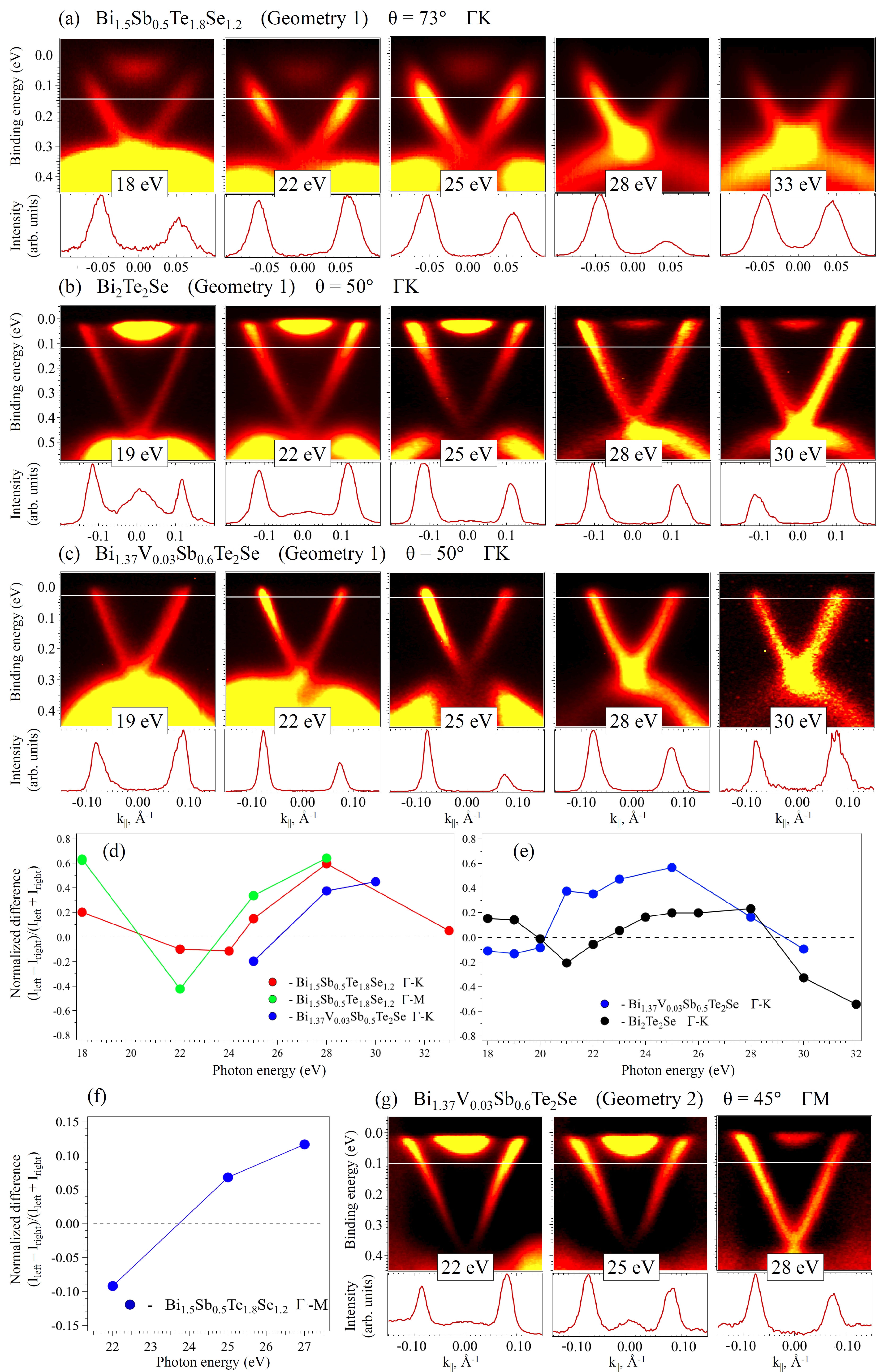}
\caption{Lines (a,b,c) -- series of ARPES intensity maps of the TSSs measured along the SR incidence plane (\GK) (Geometry \textbf{1}) for pristine TIs with stoichiometry Bi$_{1.5}$Sb$_{0.5}$Te$_{1.8}$Se$_{1.2}$, Bi$_2$Te$_2$Se and for V-doped TIs with stoichiometry Bi$_{1.37}$V$_{0.03}$Sb$_{0.6}$Te$_2$Se, accordingly, by using linear $p$-polarized SR at different photon energy. The profiles of comparable intensities of the TSSs with opposite momenta cut at the energy corresponding to enhanced intensity (marked by horizontal white lines) are presented at the bottom of each insets. (d,e) -- The TSS intensity asymmetry (A) trend vs photon energy estimated from the ARPES intensity maps presented in lines (a,b,c) and Supplementary Figs.~1S,~2S measured along \GK and \GM directions of the Brillouin zone. (g) -- Similar ARPES intensity maps measured in the direction perpendicular to the SR incidence plane (Geometry \textbf{2}) for pristine TIs with stoichiometry Bi$_{1.5}$Sb$_{0.5}$Te$_{1.8}$Se$_{1.2}$ and (f) -- corresponding variation of the TSS intensity asymmetry (A).  
\label{Fig1}}
\end{figure}

Figures~\ref{Fig1}(a,b,c) show the photon energy dependence of the TSS
ARPES energy-momentum intensity maps measured along the \GK direction
of the surface Brillouin zone using linearly $p$-polarized SR for
pristine and for magnetically-doped TIs with the following
stoichiometries: (a) Bi$_{1.37}$Sb$_{0.5}$Te$_{1.8}$ Se$_{1.2}$, (b)
Bi$_{2}$Te$_{2}$Se, and (c)
Bi$_{1.37}$V$_{0.03}$Sb$_{0.6}$Te$_2$Se. In Fig.~\ref{Fig1}(g), the
ARPES dispersions measured along the \GM direction for
Bi$_{1.37}$Sb$_{0.5}$Te$_{1.8}$Se$_{1.2}$ are shown for 
comparison. The geometry of the experiments is schematically presented
in Methods Fig.~\ref{Fig1M}.  For the presented experiments the
Geometries \textbf{1} and \textbf{2} were used with the analyzer
entrance slit orientation along and perpendicular to the SR incidence
plane. For Geometry \textbf{1} two different photon incidence angles
have been used: 73{\degree} (a) and 50{\degree} (b),(c) relative to
the surface normal (with measurements along \GK). The intensity maps
for Bi$_{1.5}$Sb$_{0.5}$Te$_{1.8}$Se$_{1.2}$ along \GM, for
Bi$_{1.4}$Sb$_{0.6}$Te$_2$Se along \GK and for
Bi$_{1.37}$V$_{0.03}$Sb$_{0.6}$Te$_{2}$Se along \GM are presented in
Supplementary Figs.~1S, 2S and 3S, respectively, with different
orientation of the SR incidence plane. For Geometry \textbf{2}, the
incidence angles was 45{\degree} (g) with measurements along \GM,
orthogonal to the SR incidence plane (oriented along \GK).  Below each
ARPES intensity map the profiles of the TSS intensities are presented 
for the constant-energy cut of the upper DC at 
the binding energy corresponding to a high intensity 
of the TSSs, see white lines in the ARPES maps.
In all the profiles the intensities at opposite $k_\parallel$ points
are different. At some photon energies the intensity at positive
$k_\parallel$ is larger than at negative $k_\parallel$, and 
at other photon energies the relation is opposite. 
Figs.~\ref{Fig1}(d),(e),(f) collect all data
measured for different TIs and demonstrate the photon energy
dependence of the asymmetry in the intensity of the opposite TSS
branches. The TSS intensity asymmetries are characterized by the ratio
\begin{equation}
\label{eq:Asym}
A = \frac{I(-k_{\parallel})-I(k_{\parallel})}{I(-k_{\parallel})+I(k_{\parallel})},
\end{equation}
measured at the energies marked by white horizontal lines. The asymmetry $A$ oscillates with photon energy both for pristine and for V-doped TIs. 

\begin{figure*}[b]
\centering
\includegraphics[width=0.7\textwidth]{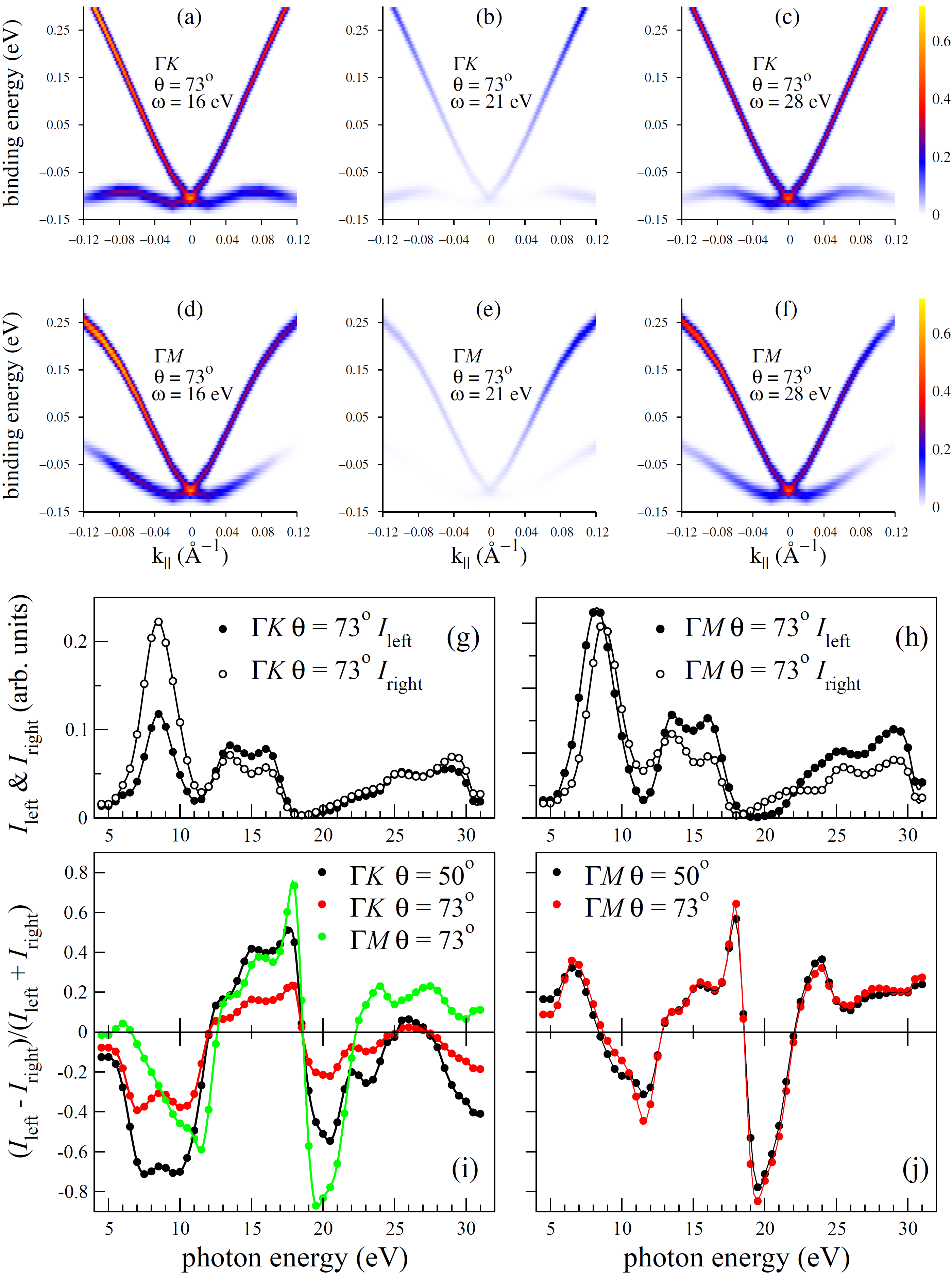} 
\caption{\label{Fig2}(a-f) Calculated energy-momentum distribution of the photoemission intensity from the DC surface 
states along $\bar\Gamma\bar K$ (a-c) and along $\bar\Gamma\bar M$ (d-f) for the three photon energies: 
$h\nu=16$~(a,d), 21~(b,e), and 18~eV (c,f). In all cases a $p$-polarized light is incident along 
$\bar\Gamma\bar K$ at an angle of $\theta=73^\circ$. (g,h) Photon energy dependence of the total 
intensity from the upper Dirac cone integrated over negative $k_\parallel$  ($I_{\rm left}$, full circles) 
and positive $k_\parallel$ ($I_{\rm right}$, open circles) for $k_\parallel$ along $\bar\Gamma\bar K$ (g) 
and along $\bar\Gamma\bar M$ (h). (i,j) Normalized difference between the integrated intensities in the 
opposite directions, $[I_{\rm left}(h\nu)-I_{\rm right}(h\nu)]/[I_{\rm left}(h\nu)+I_{\rm right}(h\nu)]$ for
two angles of incidence, $\theta=50$~and~$73^\circ$. In graph (i) the light incidence plane is parallel 
to $\mathbf{k}_\parallel$ (both for $\bar\Gamma\bar K$ and for $\bar\Gamma\bar M$), and in graph (j) it 
is perpendicular to $\mathbf{k}_\parallel$ ($\bar\Gamma\bar M$).
}
\end{figure*}

In order to explain the observed $k_\parallel$-distribution of the photocurrent and its 
variation with the photon energy we have calculated {\it ab initio} the energy-momentum 
distribution of the photoemission intensity from the DC of the stoichiometric compound Bi$_2$Te$_2$Se. The electronic structure of 
Bi$_2$Te$_2$Se \cite{Miyamoto2012} is quite similar to that of the crystals with the 
fractional stoichiometry studied here. We use the one-step theory of photoemission
in the dipole approximation, so the photocurrent from the state $\ket{{\mathbf k}_\parallel}$ 
is proportional to the transition probability $|\bra{\Phi}-i\nabla_{\mathbf e}\ket{{\mathbf k}_\parallel}|^2$ 
to the time-reversed low energy electron diffraction state $\ket{\Phi}$~\cite{Adawi64}, where 
$-i\nabla_{\mathbf e}$ is the momentum operator in the direction of the light polarization $\mathbf e$. 
The final state $\ket{\Phi}$ is calculated for the scattering of electrons on a semi-infinite 
crystal as explained in Ref.~\cite{Krasovskii99}. The inelastic scattering is included by 
adding a spatially constant imaginary part $V_{\rm i}=1$~eV to the crystal potential. The crystal 
potential was obtained within the local density approximation  with the full-potential linear 
augmented plane wave method~\cite{KSS99}. The initial states were calculated within a two-component 
relativistic formalism~\cite{KOH77} for a slab composed of 7 quintuple layers. Figs.~\ref{Fig2}(a-c) 
show the $k_\parallel$-distribution of the calculated photoemission intensity from the DC of Bi$_2$Te$_2$Se for 
the opposite \GK directions and Figs.~\ref{Fig2}(d-f) for the opposite \GM directions. In both cases
the $p$-polarized SR is incident along \GK (Geometry \textbf{1}). The $k_\parallel$-integrated 
intensities for \GK and \GM are shown in Fig.~\ref{Fig2}(g) and \ref{Fig2}(h), respectively.
The corresponding asymmetry index for the light incident along ${\mathbf k}_\parallel$ is shown in Fig.~\ref{Fig2}(i) 
and for the light incident perpendicular to ${\mathbf k}_\parallel$ (Geometry \textbf{2}) in Fig.~\ref{Fig2}(j). 

Note that when the light is incident along $\mathbf{k}_\parallel$ the difference between 
$+k_\parallel$ and $-k_\parallel$ is due to a linear dichroism, and it strongly depends on 
the angle of incidence, see Fig.~\ref{Fig2}(i). In particular, along $\bar\Gamma\bar K$ 
the opposite directions are equivalent owing to the $C_{3v}$ symmetry of the surface, and 
it is the light that breaks the symmetry of the experiment. On the contrary, for the light
incident perpendicular to $\mathbf{k}_\parallel$, Fig.~\ref{Fig2}(j), the intensity asymmetry
is due to the inequivalence of $+k_\parallel$ and $-k_\parallel$ along $\bar\Gamma\bar M$. In
that case the photon energy dependences $I_{\rm left}(h\nu)$ and $I_{\rm right}(h\nu)$ are
very similar, see Fig.~\ref{Fig2}(h), and the intensity asymmetry does not strongly depend 
on $\theta$, see Fig.~\ref{Fig2}(j). The theory explains the experimentally observed oscillations 
of the asymmetry index with the photon energy and relates them to  energy variations of the final 
state $\ket{\Phi}$.

\section*{Hole generation analysis and asymmetry variation during relaxation}

Let us assume that at the used SR energies the excited spin-polarized
photoelectrons escape to a considerable degree into the vacuum. Then,
the uncompensated spin accumulation generated by the different
photoexcitation rate of electrons with opposite momenta and its
orientation are mainly determined by the asymmetry in the concentration 
of the photo-holes in the opposite TSS branches.  How to determine this 
asymmetry? In the simplest approximation one can infer the photo-hole 
concentration asymmetry and the related uncompensated spin
accumulation from the asymmetry of the photocurrent from the DC states.

The second question is whether the asymmetry in the concentration of
the photo-holes is preserved during the relaxation process.  To
clarify this we carried out a time-resolved pump-probe laser
experiment, as in Refs.~\cite{Hajlaoui2014, Neupane2015, Crepaldi2013,
  Ishida2015}. The modification of the TSS intensity asymmetry
measured for Bi$_{1.97}$V$_{0.03}$Te$_{2.4}$Se$_{0.6}$ immediately
after the photoexcitation (just under the laser pump pulse generation)
and during the relaxation of photoexcited electrons is presented in
Fig.~\ref{Fig3}(a). The scheme and geometry of the experiment
\cite{Neupane2015, Ishida2015} are presented in Supplementary
Fig.~4S. The probe pulse of $h\nu= 5.9$~eV was linearly
$p$-polarized. The pump pulse was $s$-polarized with $h\nu$=1.48~eV.
Below each ARPES map the intensity asymmetry profiles of the
opposite DC branches are shown. These asymmetries were measured at an energy near the
bottom of the conduction band (CB) marked by the horizontal dashed
lines. These data show the time evolution of the intensity
distribution of both the photoexcited electrons (above the Fermi
level) and the de-occupied DC states (below the Fermi level). The
variation of the depletion of the TSS intensity below the Fermi level
can be an indicator of the hole generation and relaxation. One can see that 
the TSS intensity asymmetry observed at the moment of the electron photoexcitation 
is practically not changed during relaxation. 
The unchanged with time TSS asymmetry is observed also for other energy cuts, 
see Supplementary Fig.~5S. It allows us to make a very important qualitative conclusion 
that within this approximation the photo-hole generated asymmetry is not significantly transient with time. 
In other words, the asymmetry in the measured PE intensity of the TSSs with
opposite momenta can be used for a rough estimate of the induced 
uncompensated spin accumulation and the related magnetization, as we
discuss in detail in the next section.

\section*{Estimations of the in-plane and out-of-plane spin polarization (magnetization) induced in V-doped and pristine TI by linearly polarized SR}

In the following we discuss the theoretical estimations of the magnetization induced by linear $p$-polarized SR due to the generated uncompensated spin accumulation. First of all let us consider the symmetry of the spin accumulation induced by linearly polarized SR. Taking into account that this effect is caused by photoexcitation, the spin density $\dot{\vec{S}}$ can be considered as quadratic by the field:
\begin{equation}
\label{eq:Sdot}
\dot{\vec{S}} = B_{ikl} E_k E_l,
\end{equation}
where $B_{ikl}$ is a 3rd rank pseudotensor and $E_{k,l}$ are the components of electric field. Following to Refs.~\cite{DiSalvo1999, xia2009} the Bi-based compounds are referred to the space group $D^5_{3d}$($R\overline{3}m$). In the presence of a [111] surface, the symmetry
of the considered four-component complex compound can be reduced to $C_3$, and therefore $B_{ikl}$ is equal \cite{Sirotin-1979}:
\begin{equation}
\label{eq:Lam}
B_{ikl} = \begin{pmatrix} B_{11} & -B_{11} & 0 & B_{14} & B_{15} & -B_{22} \\ -B_{22} & B_{22} & 0 & B_{15} & -B_{14} & -B_{11} \\ B_{31} & B_{31} & B_{33} & 0 & 0 & 0 \end{pmatrix},
\end{equation}
where $B_{i \mu} = B_{ikl}$ $(kl \leftrightarrow \mu = 1, ... , 6)$ are the coefficients determined by the material. Considering the SR radiation as $\vec{E}= (E \cos \psi, 0, E \sin \psi)^T$, from the Eq.~\ref{eq:Lam} it immediately follows that SR can excite both the in-plane and out-of-plane uncompensated spin accumulation. Assuming that the linearly polarized SR can be decomposed on right and left circularly polarized SR, which of them can depopulate mostly one of the Dirac cone branches \cite{Shikin2016APL}, the averaged uncompensated spin accumulation can be represented empirically as $\delta S_{x,z}= \frac{\hbar}{2} \xi_{x,z} P \tau A$, where $P$ is the probability of the electron photoexcitation per unit time, $\tau$ is the decoherence time of the spins and $\xi_{x,z}$ is an empirical constant. It should be noted that $P \tau$ is the steady-state concentration of the generated holes. In the simplest case of semiconductor optical orientation $\xi_x^0 = \sin \psi$ and $\xi_z^0 = \cos \psi$ \cite{Meier-2012}. For simplicity, we assume that $\xi_{x,z} = \kappa_{x,z} \xi_{x,z}^0$, where $\kappa_{x,z} \approx 1$. Assuming that the magnetization of electron sub-system induced by linearly polarized SR can be estimated as $m_{x,z}= \mu_B \kappa_{x,z} \xi_{x,z}^0 P \tau A$, where $\mu_B$ is the Bohr magneton and $A$ is the TSS photoexcitation asymmetry (see also \cite{Shikin2016APL}). 

In the case of the V-impurity subsystem, below the Curie temperature $(T < T_C)$ the total energy can be represented as $E_V=-K_U (m_z^V)^2 - (\vec{m}^V \cdot\vec{H}_{SR})$ where $K_U$ is the constant of uniaxial anisotropy, $m_z^V$ is the z-axis component of the V-subsystem magnetization, $\vec{H}_{SR}=H_{SR} \left(\kappa_x \sin \psi , 0 ,\kappa_z \cos \psi \right)^T$ is the field acting on the V impurities from the SR induced magnetization, $H_{SR}= \frac{\tilde{a}^2}{\mu_B} J_{eV} P \tau A$, where $\tilde{a}=4.24~\dot{A}$ is the lattice constant typical of the studied TIs, $J_{eV}\approx 0.3$~eV \cite{Rosenberg-2012, Liu-2009, Zhu-2011} is the exchange constant, which describes the $(s-d)$ interaction between the TSSs and the V-ion impurities system. In case $(T > T_C)$ the energy minimization leads to the following expression for the V-subsystem magnetization:
\begin{equation}
\label{eq:mvz}
\vec{m}^V = g \mu_B S N B_S \left( g \mu_B S H_{SR} / T \right)\left(\sin\eta , 0 ,\cos\eta \right)^T,
\end{equation}
where $g\approx 2$ is the g-factor, $S=3/2$ is the spin of impurity ion, $N$ is the averaged impurity concentration, $B_S$ is the Brillouin function and $\eta$ is slightly differs from $\psi$ and $\sin \eta = \kappa_x \sin \psi / \sqrt{\kappa_x^2 \sin^2 \psi + \kappa_z^2 \cos^2 \psi}$. In case of $(T < T_C)$ we also have to consider anisotropy, but estimations proves that the anisotropy term is significantly smaller than the SR term (see Supplementary Inform. for details), therefore, Eq.~\ref{eq:mvz} is valid with a good accuracy for all temperature range. The dependence of the in-plane component of the total magnetization $\vec{M}=\vec{m}+\vec{m}^V$ (under experimental conditions $\psi = 50^{\degree}$ and $P \tau \approx 3.5\times 10^{13}$~cm$^{-2}$) on the in-plane asymmetry value ($A$) for different temperatures is represented on Fig.~\ref{Fig4}(a).

The complete electron Hamiltonian of the considered system including both electron-electron and electron-vanadium interactions in the mean field approximation can be written as:
\begin{equation}
\label{eq:He}
\hat{H}_e = \hbar V_D \left[ \vec{k} \times \vec{\sigma} \right]\vec{e}_z + \frac{\tilde{a}^2}{\mu_B} J_{eV} \left(\vec{m}^V \cdot \vec{\sigma} \right) + \frac{\tilde{a}^2}{\mu_B} U \left(\vec{m} \cdot \vec{\sigma} \right),
\end{equation}
where $V_D \sim 5.3\times 10^7~cm/s$ is the velocity taken from the TSS dispersion law, $\vec{\sigma}$ is the vector of Pauli matrices and $U \approx 0.2$~eV is the Hubbard parameter. The energy spectrum in this case has the following form:
\begin{equation}
\label{eq:E}
E= \pm\sqrt{\hbar^2 V_D^2 |\vec{k}|^2+\Delta_x^2+\Delta_z^2-2\hbar V_D k_y \Delta_x},
\end{equation}
where 
\begin{equation}
\begin{split}
\label{eq:Deltaxz}
\Delta_x = \frac{\tilde{a}^2}{\mu_B} J_{eV} m_x^V + \frac{\tilde{a}^2}{\mu_B} U m_x \\
\Delta_z = \frac{\tilde{a}^2}{\mu_B} J_{eV} m_z^V + \frac{\tilde{a}^2}{\mu_B} U m_z \\
\end{split}
\end{equation}

The resulting band structure modification under influence of the induced out-of-plane and in-plane magnetization is shown in Fig.~\ref{Fig4}(b) (under experimental conditions noted above and the averaged value of the asymmetry of $A=0.5$ taken from Fig.~\ref{Fig5}(a)). The out-of-plane magnetization is accompanied by the splitting of the DC states at the DP. The induced in-plane magnetization emerges in the $k_{\parallel}$-shift of the DC in the direction orthogonal to the induced magnetic field (or magnetization). (Analogous $k_{\parallel}$-shift is observed under external applied in-plane magnetic field~\cite{Ogawa2016, Henk-2012PRL, Semenov2012}). Corresponding calculated temperature dependence of the DP-gap value and the $k_{\parallel}$-shift of the DP position in the direction orthogonal to the magnetization are presented in Figs.~\ref{Fig4}(c,d) (see the discussion below).

\subsection*{Experimental confirmation of the induced in-plane magnetization}

As an experimental evidence of the SR-induced in-plane magnetization, 
Fig.~\ref{Fig5} demonstrates a correlation between the
 modification of the experimental intensity maps of the upper
DC states close to the Fermi level
(line~\ref{Fig5}(a)) and the ($k_{x},k_{y}$)-shift of the DP position
induced by the in-plane magnetic field generated by SR with linear
$p$- and opposite circular polarizations (line~\ref{Fig5}(b)), which is expected in accordance with
Fig.~\ref{Fig4}(b). The incident direction of SR corresponds to the vertical line. For each of the DC
energy cuts (Fig.~\ref{Fig5}(a)) the TSS intensity profiles in the
$k_{x}, k_{y}$-directions are presented, in the bottom and right graphs, respectively.
 These profiles clearly demonstrate a pronounced modulation of the intensity of the DC states intensity
and their asymmetry both along and orthogonally to the SR incidence
plane. The maps were measured with $h\nu=28$~eV for
Bi$_{1.37}$V$_{0.03}$Sb$_{0.6}$Te$_2$Se kept at the temperature of
55K. The DP positions (Fig.~\ref{Fig5}(b)) were estimated from the
maximal intensity of the TSS in $k_{x}$ and $k_{y}$, see the profiles
at the bottom and on the right side of the maps cut at the DP. The
different depopulation of the opposite DC states under photoexcitation
with different polarization of SR is well visible in Fig.~\ref{Fig5}(a), and it 
leads to the ($k_{x},k_{y}$)-shift of the DC according to the direction of induced magnetic field,
which is determined by the asymmetry in the TSS intensity. The
direction of the uncompensated spin accumulation and of the
induced in-plane magnetic field are determined by the direction where
the TSS intensity asymmetry is oriented and by the details of the
spin texture \cite{Zhang2013PRL, Cao2013, Zhu2013}. The relation
between the experimental TSS intensity asymmetry (A), the direction of
the induced magnetic field (M) generated by uncompensated spin
accumulation (S) in correspondence to spin texture and direction of the DP position 
($k_{x}, k_{y}$)-shift is schematically shown in
Fig.~\ref{Fig5}(c). For circularly polarized SR a pronounced asymmetry
in the TSS intensity is observed in the direction perpendicular to the
SR incidence. The use of the opposite circular polarizations leads to the TSS intensity asymmetry and generated uncompensated 
spin accumulation with spin orientation (and corresponding magnetic moment) 
in opposite $k_{\parallel}$-directions. This direction of the 
induced magnetic field determines the shift of the DP orthogonally to the SR incidence
($k_{x}$) that is confirmed experimentally (see the shift of blue
crosses in comparison with the green one). For linear $p$-polarization
of SR the TSS intensity asymmetry is observed in the direction along
the SR incidence. This leads to a spin accumulation (S) and magnetic field (M)
perpendicular to the SR incidence. As a result, the $k_y$-shift of the
DP position is observed in the direction orthogonal to that in the
case of a circularly polarized SR. The measurements at room
temperature (Supplementary Fig.~6S) and at 30K at the 9B beamline
HiSOR (Hiroshima, Japan) albeit with a lower intensity of SR (not
shown) demonstrate similar behavior. The value of the
$k_{\parallel}$-shift of the Dirac cone induced by linearly-polarized
SR shown in Fig.~\ref{Fig5} (in comparison with the DC position under
excitation by circularly polarized SR) can be estimated to approximately 
$5-10\times10^{-3}$ \AA$^{-1}$ in agreement with the
theoretical estimation of the $k_{\parallel}$-shift in
Fig.~\ref{Fig4}(b). Lower value of the $k_{\parallel}$-shift of the DP
position at room temperature presented in Supplementary Fig.~6S
confirms the calculated temperature dependence of the DP
$k_{\parallel}$-shift shown in Fig.~\ref{Fig4}(d). Under excitation
by a circularly polarized SR of opposite chirality the
$k_{\parallel}$-shift of the DC position is stronger, which is related
to the enhanced TSS intensity asymmetry (see corresponding profiles).

Additionally, the in-plane magnetization induced by a linearly
polarized SR can be confirmed by the $k_{\parallel}$-shift of the
spin-polarized DC states relative to the non-spin-polarized CB states. 
(A similar $k_{\parallel}$-shift under applied
magnetic field was noted in Ref.~\cite{Ogawa2016}). Supplementary Fig.~7S demonstrates 
that such $k_{\parallel}$-shift is actually observed at different photon energies, and 
it is important that the direction of the 
$k_{\parallel}$-shift is related to the observed TSS intensity asymmetry. It is interesting 
that the value of the $k_{\parallel}$-shift relative to $k_{\parallel}=0$
can be also estimated to be about $10\times10^{-3}$ \AA$^{-1}$, which correlates with the estimations noted above.
Moreover, when the sign of the asymmetry changes with photon energy the DC shifts in the opposite direction. 
This confirms that the observed shift is really connected
with the TSS intensity asymmetry and the resulting induced in-plane magnetization direction. The $k_{\parallel}$-shift of the DC branches relative the CB states located at $k_{\parallel}=0$ is also observed in the case of photoexcitation by laser radiation. In that case the profiles shown in Fig.~\ref{Fig3}(b) below the ARPES dispersion map
(cut at the energy marked by white line which crosses both DC and CB
states) demonstrate non-equal distance from the left and right side.

As a partial conclusion we have experimentally observed the DP position ($k_{x},k_{y}$)-shifts in accord with the measured asymmetry of TSS intensity confirming that the non-equal depopulation of these states induces the in-plane magnetic fields under photoexcitation by SR.

\subsection*{Out-of-plane induced magnetization and its experimental confirmation}

Let us now discuss the problem of the out-of-plane magnetization induced by linearly polarized SR. 
Fig.~\ref{Fig6}(a) shows the calculated photon energy dependence of the out-of-plane net-spin 
photocurrent $S(k_{\parallel},h\nu)=I^{\uparrow}(k_{\parallel},h\nu)-I^{\downarrow}(k_{\parallel},h\nu)$ 
from the upper DC of Bi$_2$Te$_2$Se with the SR incident along \GK. The out-of-plane 
net-spin-photocurrent integrated over $k_{\parallel}$ does not vanish, and its photon energy
dependence is shown in Fig.~\ref{Fig6}(b). 
The magnitude and the sign of the integral photocurrent are seen to
vary with the photon energy, which suggests that also the total
out-of-plane spin accumulation may be different for different photon
energies. This out-of-plane spin accumulation generates the induced
out-of-plane magnetization that should lift 
the degeneracy of the TSSs and open the energy gap at the DP
due to the TRS breaking, as in
Refs.~\cite{Shikin2016APL, Shikin-2017_2D} under circularly polarized SR.
Using Eq.~\ref{eq:Deltaxz} one can estimate the energy gap as $\Delta = 2\frac{\tilde{a}^2}{\mu_B}
J_{eV} m_z^V + 2\frac{\tilde{a}^2}{\mu_B} U m_z$. 

Fig.~\ref{Fig4}(c) shows the calculated temperature dependence of the
gap induced at the Dirac point using the value of the TSS asymmetry of
0.3 (as averaged value taken from the asymmetry values presented in Fig.~\ref{Fig5}(a) 
and Fig.~\ref{Fig1}(e) at $h\nu=28-30$~eV) both for magnetically-doped and for pristine TIs at
the experimental conditions noted above (see Methods for details).

For magnetically-doped TI below 30~K a gap of about 25~meV is expected. As temperature increases the gap decreases down to $12-15$~meV at room temperature. A finite gap above the Curie temperature is just related to the out-of-plane magnetization generated by SR. In the case of pristine TI a formation of the gap of 4.5~meV is expected too, independently on temperature. Unfortunately the energy resolution in our conditions was not high enough to allow the measurement of this small gap value within a reasonable experimental error.

The experimentally measured in-plane and out-of-plane spin polarization is presented 
in spin-resolved spectra in Figs.~\ref{Fig7}(a,b). The spectra were measured for
Bi$_{1.31}$V$_{0.03}$Sb$_{0.66}$Te$_2$Se at the temperature of 23~K at the DP by using 
linear $p$-polarized SR. The corresponding experimental polarization asymmetry is plotted in the bottom part of the graphs. 
The spin orientation was measured along the SR incidence 
plane (Geometry \textbf{1}). The related ARPES intensity map is shown in the inset 
in the upper central part. The spin-resolved spectrum in Fig.~\ref{Fig7}(a) 
confirms the in-plane spin-polarization of the states in the region of 
the DP, which is inverted relative to the DP (see 
black arrows on the polarization asymmetry). The observed
in-plane polarization along the SR incidence plane is determined by
the contribution of the $p_x$, $p_y$ components in the spin
texture. Similar effects are described in Refs.~\cite{Schmidt-2011, Seibel2016}. The generated out-of-plane polarization is shown in
Fig.~\ref{Fig7}(b). The presented spin-resolved spectrum demonstrates
availability of the spin-polarized states at the Fermi level. We
ascribe these states to the V 3$d$-resonances characterized by the
out-of-plane spin polarization. It is known that for V-doped TIs the
maximal intensity of the V 3$d$-ion states is located near the Fermi
level \cite{Vergniory2014, Yu2010}. The magnetically-doped TIs are
characterized by a colossal anisotropy of the magnetic moment induced
at the FM-impurity atoms, which favors the magnetization perpendicular
to the surface. Therefore, the observed out-of-plane spin polarization
of the states near the Fermi level can signify the out-of-plane magnetization of the V 3$d$-ions.

The SR-induced out-of-plane magnetization can be also experimentally
confirmed by the splitting of the TSSs and opening of the energy gap
at the DP both below and above the Curie temperature (see discussion
above). Figs.~\ref{Fig7}(c,d) demonstrate the spin-integrated spectra
measured directly at the DP for V-doped TIs at the temperatures of 1~K
and 66~K, below and above the Curie temperature, respectively (which
is below 5-10~K), in comparison with that measured for pristine TI at
room temperature -- Fig.~\ref{Fig7}(e). The relevant ARPES intensity
maps are presented above each spectrum to show the DP positions. The
fitting procedures for the spectra measured for V-doped TIs
(Figs.~\ref{Fig7}(c,d)) show a decomposition into two spectral components with the energy splitting of about 20-25~meV at
the DP below and above the Curie temperature. The width of the
components was chosen in accord with the width of the TSS peaks
outside the DP extracted from the ARPES intensity maps.
The splitting into two components for the states at the DP above the
Curie temperature (when a spontaneous magnetic order is destroyed)
allows us to conclude that the out-of-plane magnetization is induced
by SR. The spectra for other TIs presented in Supplementary Fig.~8S
demonstrate similar behavior with the splitting of the TSSs at the
DP. At the same time, the spectra measured at the DP for pristine TI
(Fig.~\ref{Fig7}(e) and Supplementary Fig.~8S) do not show a
noticeable splitting at the DP. The fitting procedure for these
spectra shows only a one-component structure.  The splitting of
4.5~meV predicted by theoretical estimations for pristine TI
(Fig.~\ref{Fig4}(b)), is not resolved in our experiment. Thus, for TI
without magnetic doping the photoexcitation by SR does not lead to a noticeable
 out-of-plane magnetization. Therefore, one can
conclude that the splitting of the states at the DP in magnetically
doped TIs can actually be an indicator of the induced out-of-plane
magnetization generated by the linearly-polarized SR.

\textbf{In summary}, we have demonstrated that the asymmetry of the DC
states with opposite spin orientation in photoexcitation by
$p$-polarized SR is accompanied in pristine and magnetically-doped TIs
by an uncompensated spin accumulation of the generated holes in
the initial states. This leads to the in-plane and out-of-plane polarization of the TSSs and the V 3$d$-ions and
corresponding magnetization via the spin-torque effect. Experimentally
it is indicated by spin-resolved PE spectra and is confirmed by the
$k_{\parallel}$-shift of the DC position under the induced in-plane
magnetic field and by the splitting of the TSSs at the DP induced by
the out-of-plane component of magnetization even above the Curie
temperature. The laser pump-probe experiment has shown that the
difference in depopulation of the opposite branches of the DC states
is practically not changed during the relaxation process. It allows to
conclude that the SR-induced magnetization can be roughly estimated
using the asymmetry in the intensity of the TSSs in PE
spectra. Theoretical estimations have confirmed a possibility of the
induced in-plane and out-of-plane magnetization by linearly polarized
SR.

This finding should be taken into account in PE investigations of systems with helical spin structure, especially for magnetically-doped TIs, where the PE process can influences the spin structure in the ground state depending on photon energy and experimental details. 

\section*{Methods}
The measurements of ARPES intensity maps for the DC states in pristine
and magnetically-doped TIs presented in Fig.~\ref{Fig1} and Supplementary Figs.~1S--3S
were carried out at i3 beamline at MAXlab (Lund, Sweden) and BaDEIPh beamline 
at Elettra (Trieste, Italy) in the direction along the SR incidence plane (Geometry \textbf{1} in Fig.~\ref{Fig1M}) 
and $1^2$ end station at BESSY II (Helmholtz-Zentrum
Berlin, Germany) in the direction perpendicular to the SR incidence
plane (Geometry \textbf{2} in Fig.~\ref{Fig1M}) using a Scienta R4000 or SPECS Phoibos 150 analyzers. 
The incidence angle of SR for these experiments was 73{\degree} (MAXlab) and 50{\degree} (Elettra and BESSY II) 
relative to the surface normal.

The spin-resolved photoemission spectra for V-doped TIs were measured
at the COPHEE setup at Swiss Light Source, Switzerland (Figs.~\ref{Fig7}(a,b)) 
and at the i3 beamline at MAXlab, Sweden (Supplementary Fig.~8S) with the spin 
orientation along the plane of the SR. The spin-resolved photoemission spectra 
were measured both for the in-plane and for the out-of-plane spin orientation. 
To increase the intensity of the TSSs in the region of the DP we used a photon energy range 
of 28-30 eV. For other photon energies the relative contribution 
of the DC states in the region of the DP is reduced. The SR incidence angle for these experiments was 
45{\degree} and 73{\degree}.

The DC energy cut maps of the TSSs for Bi$_{1.37}$V$_{0.03}$Sb$_{0.6}$Te$_2$Se (Fig.~\ref{Fig5} and
Supplementary Fig.~6S) were measured at $1^2$ station at BESSY II (Helmholtz-Zentrum Berlin, Germany) 
with a photon energy of 28~eV keeping the sample at the temperature of 55K and room temperature, respectively. 
The SR incidence angle was 50{\degree}.

The ARPES dispersion maps for magnetically-doped TIs, which were used
for careful estimation of the splitting of the TSSs at the DP
(Fig.~\ref{Fig7}) were measured at i3 beamline at MAXlab (Lund,
Sweden), at the 9B beamline at HiSOR (Hiroshima, Japan) in the
direction along the SR incidence plane (Geometry \textbf{1}) and at
$1^3$ station at BESSY II (Helmholtz-Zentrum Berlin, Germany) in the
direction perpendicular the plane of the SR incidence (Geometry
\textbf{2}). The SR incidence angle was 45{\degree} relative to the surface normal.

The time-resolved pump-probe laser experiment (Fig.~\ref{Fig3}) was
carried out in ISSP at Tokyo University (Japan) for V-doped TI with
stoichiometry Bi$_{1.97}$V$_{0.03}$Te$_{2.4}$Se$_{0.6}$. Time-resolved photoemission apparatus achieving sub-20-meV energy
resolution and high stability was used, see details in Ref.~\cite{Ishida_2014}. The probe pulse was linearly $p$-polarized and with a photon energy of 5.9 eV. The pump pulse was $s$-polarized with $h\nu$=1.48 eV. Geometry of the
experiment is presented in Supplementary Fig.~4S. The laser beam incidence angle was 45{\degree} relative to the surface normal.

Part of work was carried out in the resource center ``Physical methods of surface investigation'' (PMSI) of Research park of Saint Petersburg State University.

The single crystals of pristine TIs Bi$_{1.5}$Sb$_{0.5}$Te$_{1.8}$Se$_{1.2}$, 
Bi$_2$Te$_2$Se and Bi$_{1.4}$Sb$_{0.6}$Te$_2$Se, and magnetically-doped TIs Bi$_{1.37}$V$_{0.03}$Sb$_{0.6}$Te$_2$Se, 
Bi$_{1.3}$V$_{0.04}$Sb$_{0.66}$Te$_2$Se, and Bi$_{1.97}$V$_{0.03}$Te$_{2.4}$Se$_{0.6}$ were synthesized by using a modified vertical Bridgman method \cite{Kokh2014}. Clean surfaces of the TIs were obtained by a cleavage in ultrahigh vacuum. The base pressure during the experiments was better than $1\times10^{-10}$ mbar.

\section*{Acknowledgements}{The authors acknowledge support by the Saint Petersburg State University (grant No. 15.61.202.2015), Russian Science Foundation grants No. 17-12-01333 (in the part of theoretical study of magnetic properties and ARPES analysis) and Russian Science Foundation grant No. 17-12-01047 (in part of crystal growth and the sample characterization). The work was also supported by the Spanish Ministry of Economy and Competitiveness MINECO (Project No. FIS2016-76617-P), German-Russian Interdisciplinary Science Center (G-RISC) funded by the German Federal Foreign Office via the German Academic Exchange Service (DAAD) and Russian-German laboratory at BESSY II (Helmholtz-Zentrum Berlin). The authors kindly acknowledge the MAXLab, HiSOR, SLS, Elettra, BESSY II, ISSP of the University of Tokyo and PMSI staff for technical support and help with experiment and useful discussions.}

\section*{Author contributions}
The project was proposed by A.M.S, A.K.Z. The ARPES measurements were performed by I.I.K., D.M.S., V.Y.V., D.A.E., A.A.R., L.P., G.D.S., A.K., O.E.T., A.G.R. and A.M.S. The time-resolved ARPES measurements were performed by V.Y.V., A.K., Y.I., A.G.R. and A.M.S. Samples were synthesized and characterized by O.E.T., K.A.K. The experimental data analysis was carried out by A.M.S., A.A.R., D.M.S., V.Y.V., D.A.E. and I.I.K. Theoretical estimations and analysis of magnetic properties were performed by P.N.S., K.A.Z., A.K.Z. Calculations of PE spectra, ARPES dispersion maps and the TSS intensity asymmetry at different photon energies were performed and analysed by E.E.K., E.V.C. All authors extensively discussed the results and participated in the manuscript editing. The manuscript was written by A.M.S., E.E.K., P.N.S. and A.K.Z.

\section*{Additional information}
The authors declare that the data supporting the findings of this study are available within the paper and its supplementary information files. Supplementary information is available in the online version of the paper. Reprints and permissions information is available online at www.nature.com/reprints.

\section*{Competing financial interests}
The authors declare no competing financial interests.

\begin{figure}[ht]
\centering
\includegraphics[width=\textwidth]{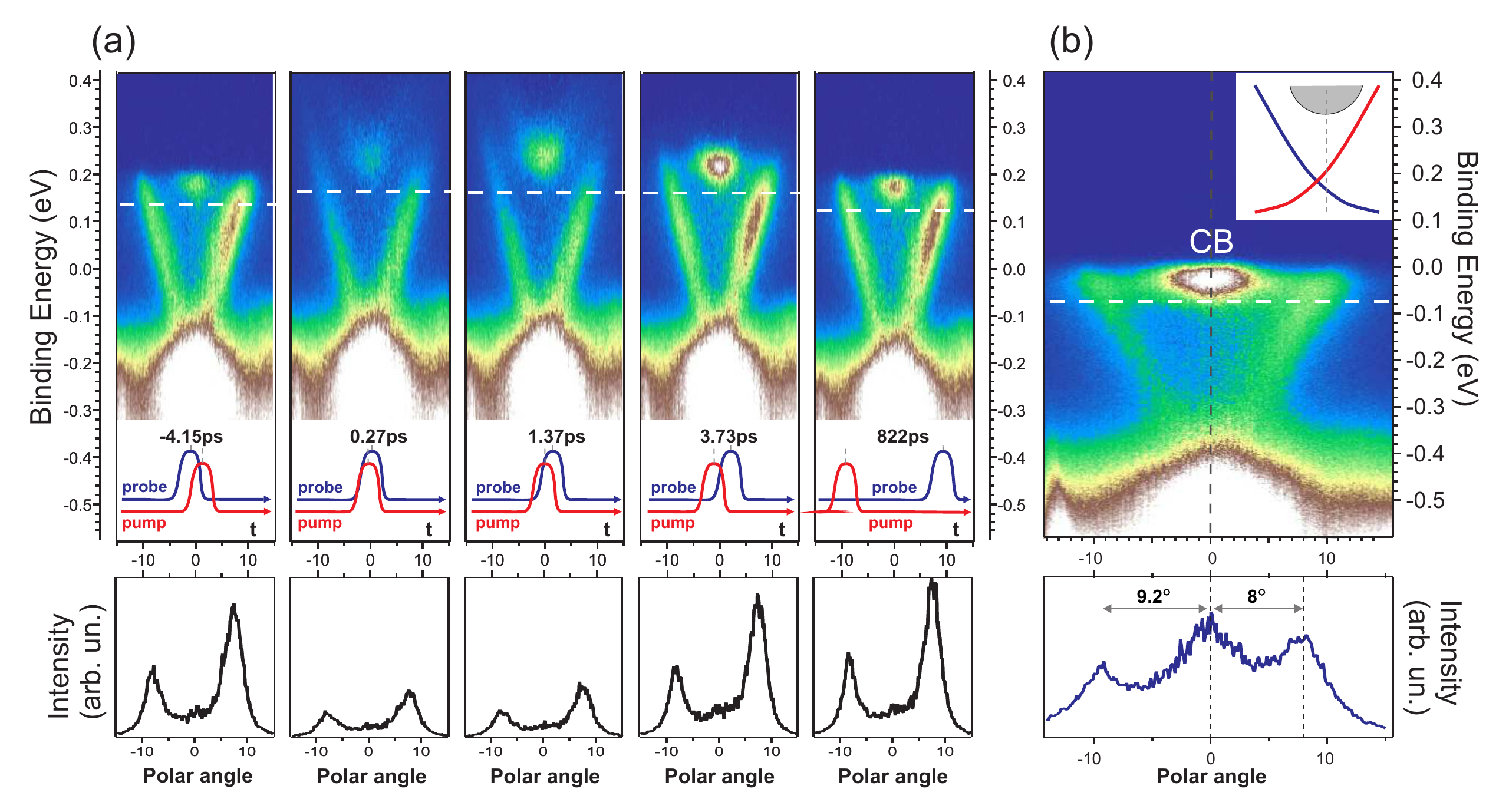}
\caption{(a) -- Time-resolved ARPES maps measured by laser pump-probe experiment for Bi$_{1.97}$V$_{0.03}$Te$_{2.4}$Se$_{0.6}$ at 11~K by using $p$-polarized probe pulse ($h\nu$=5.9 eV) and $s$-polarized pump pulse ($h\nu$=1.48 eV). The delay time between pump and probe pulses is shown below the ARPES maps. In the bottom line the time-resolved TSS intensity profiles cut at the energy close to the CB states is shown. (b) -- $k_{\parallel}$-shift of the ARPES dispersion map measured by using $p$-polarized probe pulse relative to the CB states located at the Fermi level at $k_{\parallel}=0$. Schematic presentation of the $k_{\parallel}$-shift is shown in inset. Below the ARPES map the corresponding profile of the intensities of the TSSs and CB states confirming the $k_{\parallel}$-shift of the DC states due to induced in-plane magnetic field is shown. 
\label{Fig3}}
\end{figure}

\begin{figure}[ht]
\centering
\includegraphics[width=0.9\textwidth]{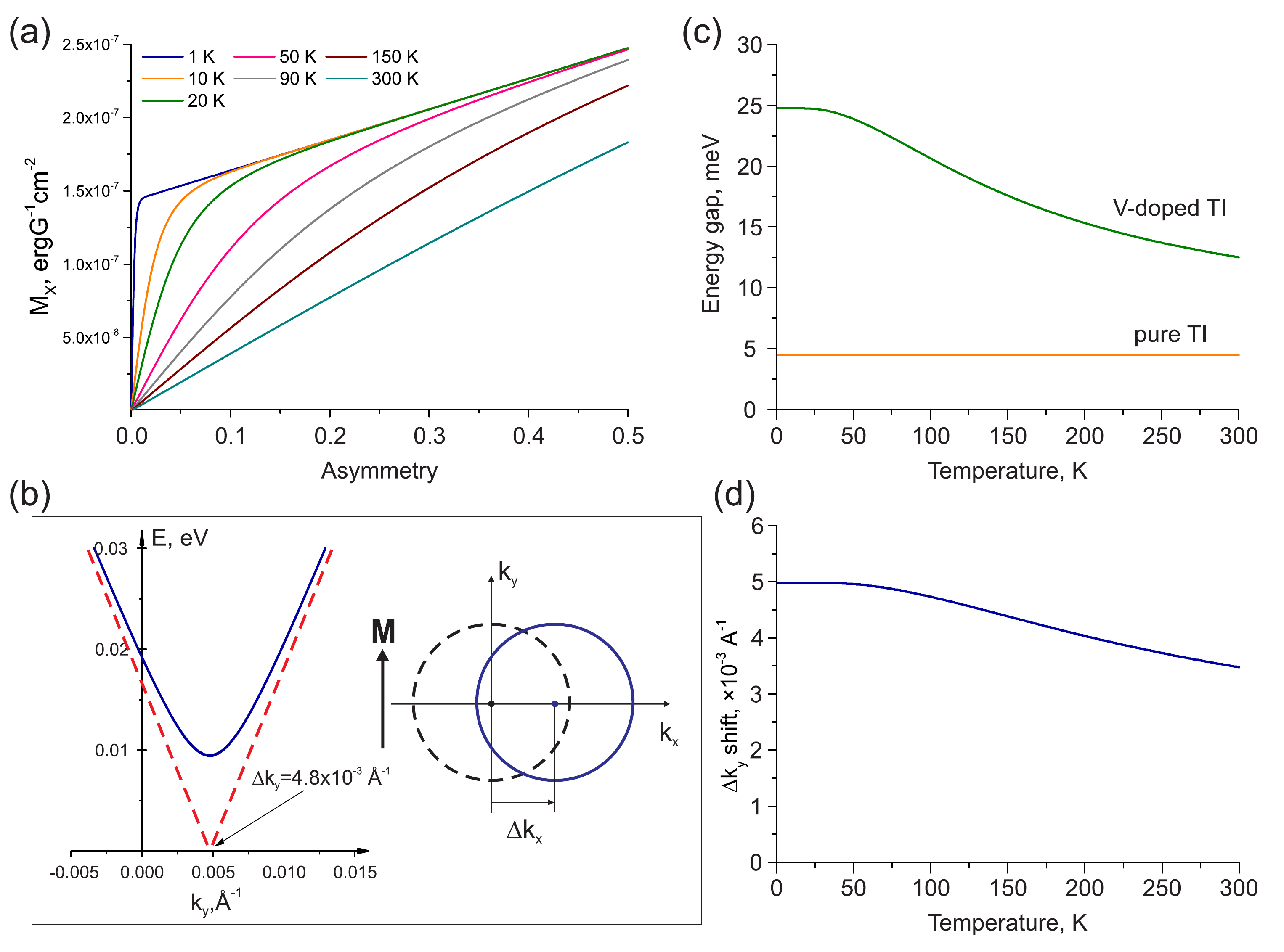}
\caption{(a) -- Calculated dependence of the in-plane component of a total SR-induced magnetization vs the asymmetry in the hole-generation with opposite spins (A) for different temperatures between 1 and 300~K. (b) -- Modification of the DC state dispersions obtained at crossing by the plane ($k_{x}$=0) which indicate the influence of the out-of-plane and in-plane components of induced magnetization (opening the gap at the DP and the $k_{x}$-shift of the DC states). (c,d) -- Calculated temperature dependences of the energy gap at the DP due to the out-of-plane component of the induced magnetization for magnetically-doped and pristine TIs and corresponding $k_{\parallel}$-shift of the DP relative to $k_{\parallel}=0$ in the direction orthogonal to the magnetization as a function of temperature.
\label{Fig4}}
\end{figure}

\begin{figure}[ht]
\centering
\includegraphics[width=0.75\textwidth]{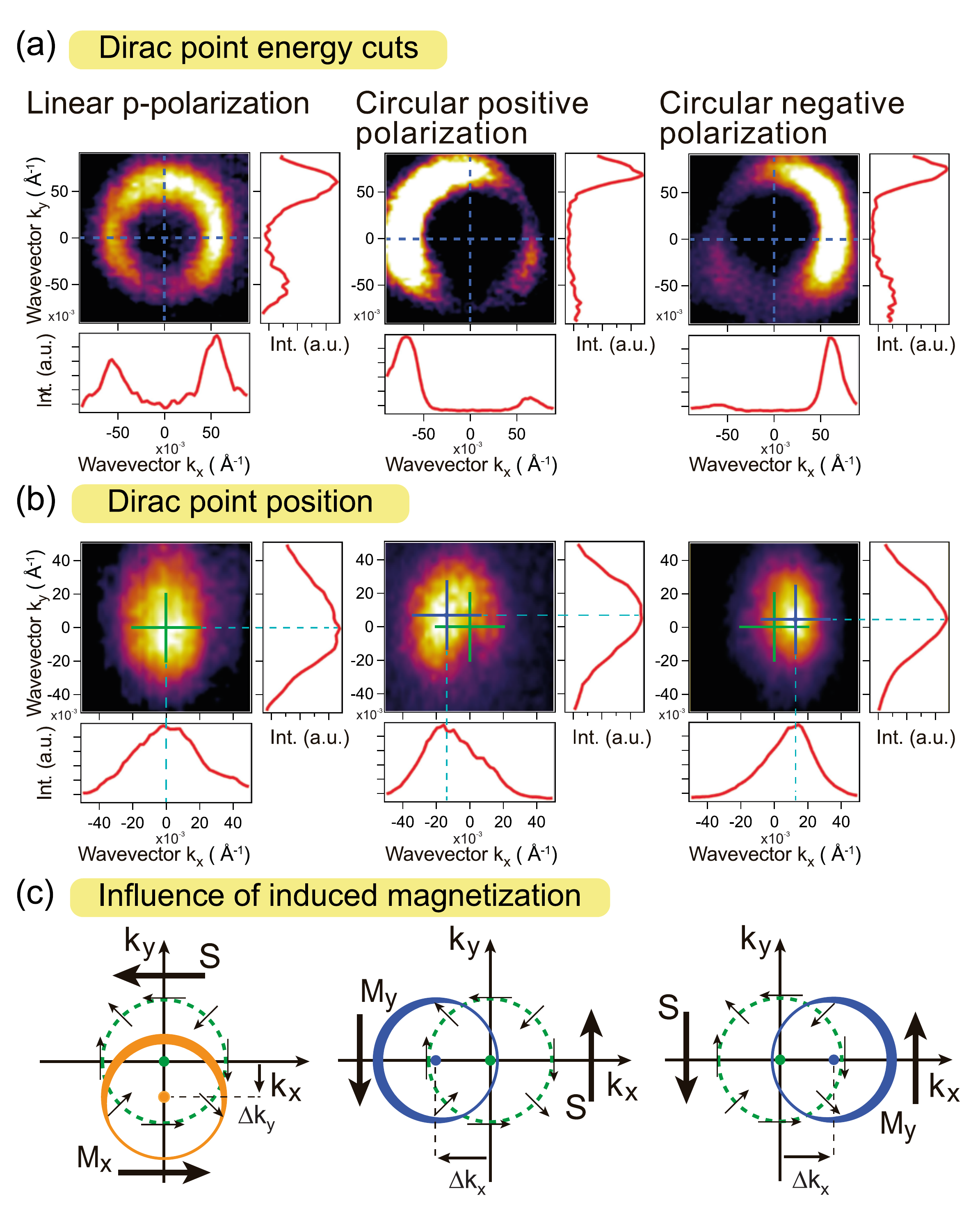}
\caption{The ($k_{x}, k_{y}$)-projections of the DC state maps cut at the energy near the Fermi level -- (a) and at the DP -- (b) measured at 55~K for Bi$_{1.37}$V$_{0.03}$Sb$_{0.6}$Te$_2$Se with different polarization of SR. Below and on the right side of the maps the corresponding profiles of the TSS intensity showing the distribution of the TSS intensity along $k_{x}$ and $k_{y}$ are presented. A pronounced asymmetry in the TSS intensity which is a source for the corresponding uncompensated spin accumulation is visible for different polarization of SR. Line (b) shows the ($k_{x}, k_{y}$)-shift of the DP positions using linearly $p$-polarized SR (marked by green crosses) and opposite circularly-polarized SR (blue crosses) under induced in-plane magnetization generated by SR. Line (c) -- schematic presentations of the relation of the asymmetry in the TSS intensity, the directions of the induced uncompensated spin accumulation (S) and the in-plane magnetization (M) and the directions of the ($k_{x}, k_{y}$)-shift of the DC position due to induced magnetization.
\label{Fig5}}
\end{figure}

\begin{figure*}[b] 
\centering
\includegraphics[width=0.8\textwidth]{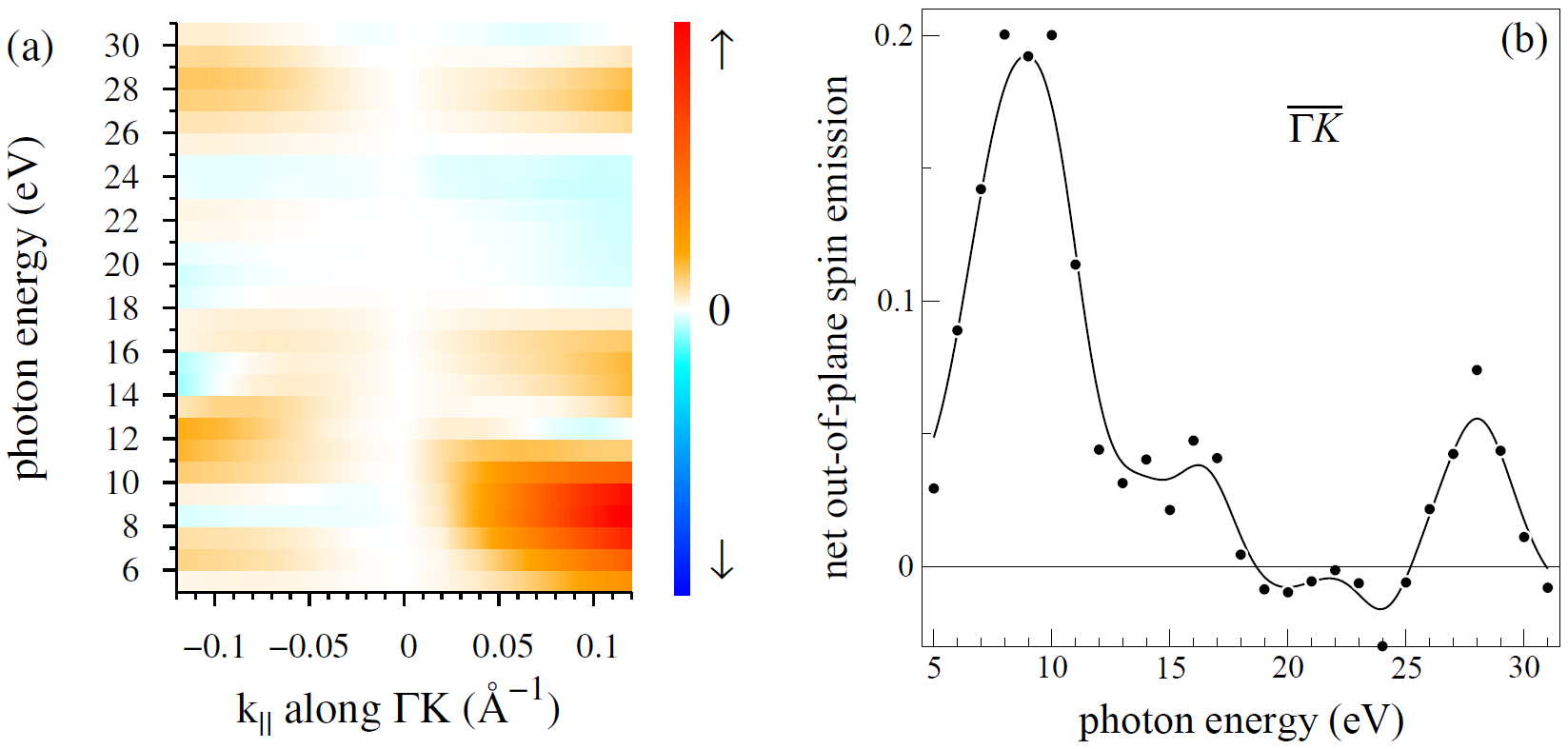} 
\caption{\label{Fig6}(a) Calculated photon-energy dependence of the $k_{\parallel}$-distribution $S(k_{\parallel},h\nu)$ 
of the net out-of-plane spin in photoelectrons excited by a $p$-polarized SR along the SR incidence 
plane (\GK) from the upper DC of Bi$_2$Te$_2$Se for $h\nu$ from 5 to 31~eV. (b) Photon energy 
dependence of the net out-of-plane spin integrated over $k_{\parallel}$ from $-0.12$ to 0.12~\AA$^{-1}$ 
along \GK.
}
\end{figure*} 

\begin{figure}[ht]
\centering
\includegraphics[width=\textwidth]{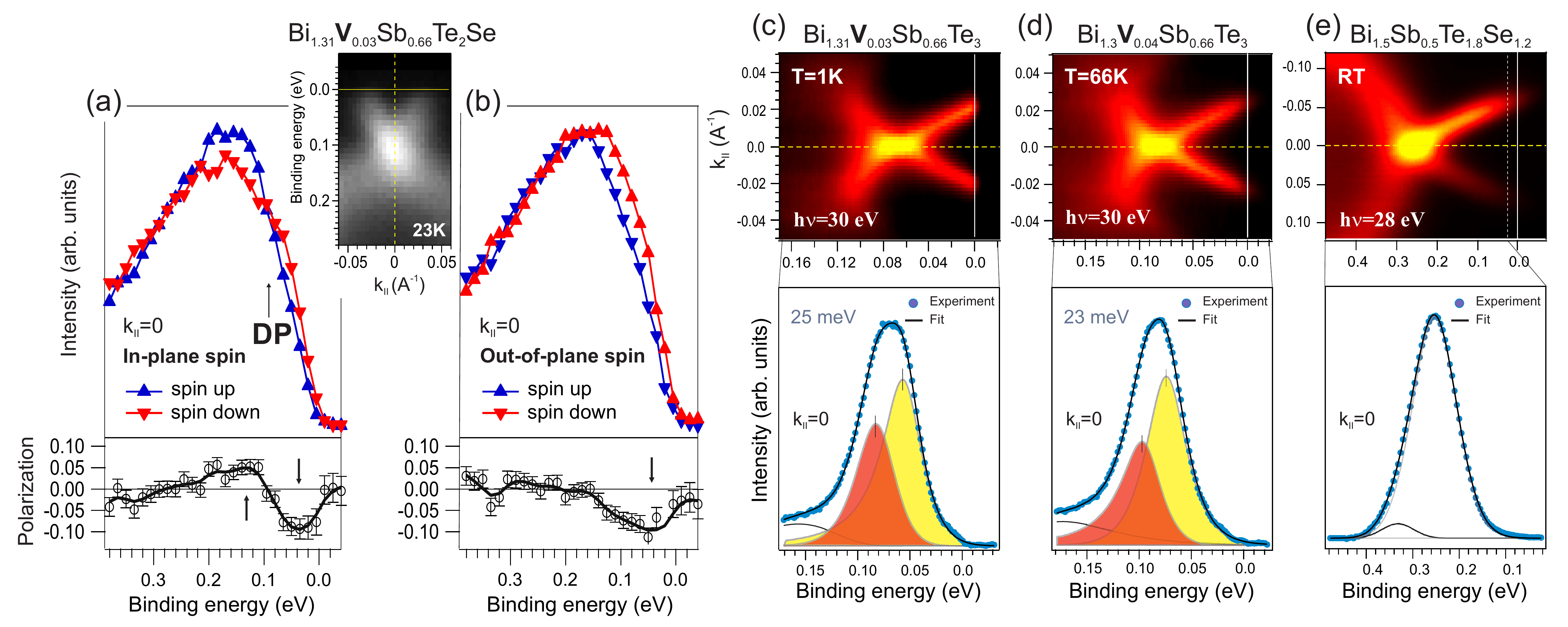}
\caption{(a,b) -- In-plane and out-of-plane spin-resolved PE spectra measured at the DP for Bi$_{1.31}$V$_{0.03}$Sb$_{0.66}$Te$_2$Se at a temperature of 23 K and a photon energy $h\nu$ = 30 eV using linear $p$-polarized SR. Corresponding experimental asymmetries for opposite spin polarization including the experimental uncertainty for each point of the measurements are shown below the spin-resolved spectra. (c-e): Upper -- ARPES intensity maps measured for V-doped -- (c,d) and pristine -- (e) TIs at different temperatures and photon energies 28 and 30 eV. Bottom -- corresponding spectra measured at the DP and the result of the fitting procedure with decomposition on spectral components.
\label{Fig7}}
\end{figure}

\begin{figure}[ht]
\centering
\renewcommand{\figurename}{\textbf{Figure 1M}}
\includegraphics[width=0.8\textwidth]{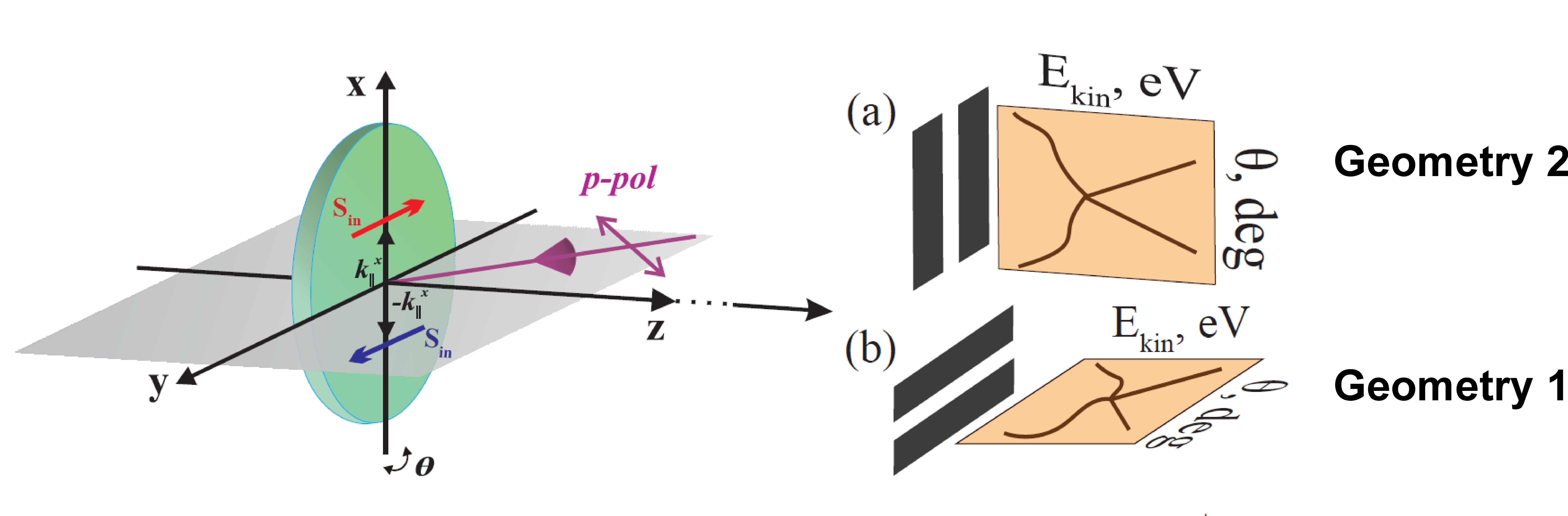}
\caption{\textbf{METHODS Section.} Schematic presentation of the ARPES measurement geometries with the analyzer slit oriented along (Geometry \textbf{1}) and perpendicular (Geometry \textbf{2}) to the of SR incidence plane.
\label{Fig1M}}
\end{figure}

\end{document}